\input{aipcheck}

\documentclass[
    ,final            
  ]
  {aipproc}

\layoutstyle{6x9}

\def\simgt{\lower 2pt \hbox{$\, \buildrel {\scriptstyle >}\over {\scriptstyle \sim}\,$}}
\def\simlt{\lower 2pt \hbox{$\, \buildrel {\scriptstyle <}\over {\scriptstyle \sim}\,$}}

\def\chandra{{\it Chandra\/}}

\def\ixo{{\it IXO\/}}

\def\suzaku{{\it {\it Suzaku}\/}}

\def\xmm{{\it XMM-Newton\/}}

\def\aox{$\alpha_{\rm ox}$}
\def\daox{$\Delta\alpha_{\rm ox}$}

\begin{document}

\title{X-raying the Winds of Luminous Active Galaxies}


\author{W.N.~Brandt}{
  address={Department of Astronomy and Astrophysics, The Pennsylvania 
State University, 525 Davey Lab, University Park, PA 16802, USA}
}

\author{G.~Chartas}{
  address={Department of Astronomy and Astrophysics, The Pennsylvania 
State University, 525 Davey Lab, University Park, PA 16802, USA},
  altaddress={Department of Physics and Astronomy, College of Charleston, 
Charleston, SC 29424, USA}
}

\author{S.C.~Gallagher}{
  address={Department of Physics and Astronomy, University of Western 
Ontario, 1151 Richmond Street, London, ON N6A 3K7, Canada}
}

\author{R.R.~Gibson}{
  address={Department of Astronomy, University of Washington, 
Box 351580, Seattle, WA 98195, USA}
}

\author{B.P.~Miller}{
  address={Department of Astronomy and Astrophysics, The Pennsylvania 
State University, 525 Davey Lab, University Park, PA 16802, USA}, 
  altaddress={Department of Physics, The College of Wooster, Wooster, OH 
44691, USA} 
}


\begin{abstract}
We briefly describe some recent observational results, mainly at \hbox{X-ray} 
wavelengths, on the winds of luminous active galactic nuclei (AGNs). 
These winds likely play a significant role in galaxy feedback. Topics 
covered include 
(1) Relations between \hbox{X-ray} and UV absorption in Broad Absorption Line 
(BAL) and mini-BAL quasars; 
(2) \hbox{X-ray} absorption in radio-loud BAL quasars; and 
(3) Evidence for relativistic iron~K BALs in the \hbox{X-ray} spectra 
of a few bright quasars. 
We also mention some key outstanding problems and prospects for future 
advances; e.g., with the {\it International X-ray Observatory (IXO)\/}. 
\end{abstract}

\maketitle


\section{The Range of AGN Outflows and the Relevance of X-ray Absorption Studies}

Fast winds that absorb \hbox{X-ray} and UV radiation are commonly seen 
in active galactic nuclei (AGNs) spanning a range of $\approx 100,000$ in luminosity. 
These outflows are a substantial component of AGN nuclear environments, and 
their ubiquity argues that mass ejection in a wind is fundamentally 
linked to mass accretion. 
In luminous AGNs, such as Broad Absorption Line (BAL) and mini-BAL quasars, 
wind velocities can be very high (up to $\approx 0.15c$), and these
winds may carry a significant fraction of the accretion power and be 
responsible for effective feedback into AGN host galaxies. In this 
proceedings paper, we will briefly review a few recent results on \hbox{X-ray} 
studies of BAL and mini-BAL quasars and also discuss some prospects for 
future advances. 

One frequently used and well-motivated model proposes that the UV 
absorption lines of BAL and mini-BAL quasars originate in a line-driven 
wind that is launched from the accretion disk at $\approx 10^{16}$~cm from 
the black hole (e.g., Murray et~al. 1995; Proga \& Kallman 2004). In 
this model, the apparently weak \hbox{X-ray} emission often seen from 
these quasars is caused by \hbox{X-ray} absorption 
(with \hbox{$N_{\rm H}\approx 10^{22}$--$10^{24}$~cm$^{-2}$})
in highly ionized ``shielding gas'' at smaller radii that protects 
the wind from nuclear EUV and soft \hbox{X-ray} radiation. Without 
this critical absorbing layer, these energetic photons would 
over-ionize the UV-absorbing gas so that it could not be radiatively 
accelerated effectively.

\section{Some Recent X-ray Results on Luminous AGN Winds}

\subsection{Statistical X-ray Studies of BAL and Mini-BAL Quasars}

\begin{figure}
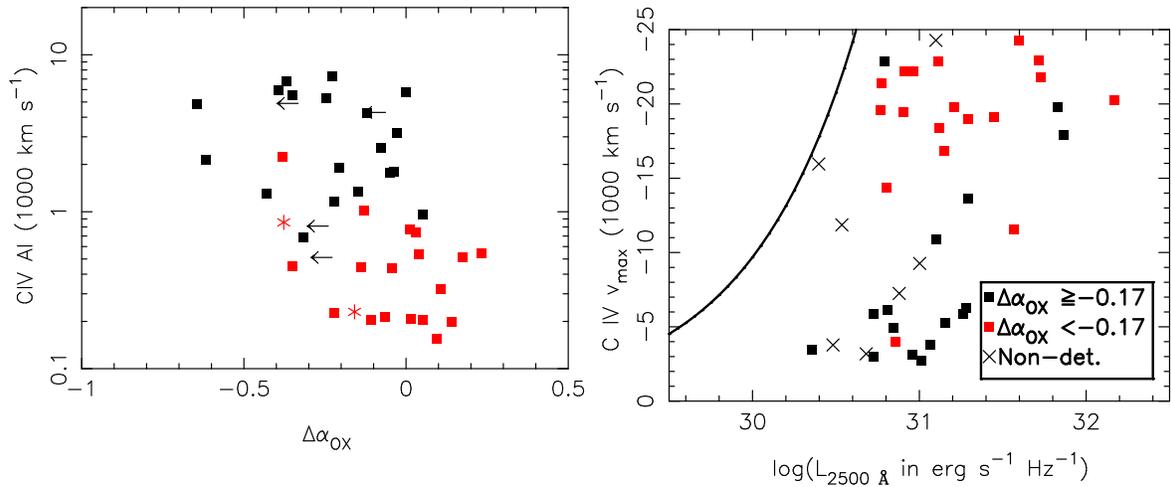

  \includegraphics[height=.35\textheight,angle=-90]{brandt_w_fig1a_color.ps}
  \includegraphics[height=.35\textheight,angle=-90]{brandt_w_fig1b_color.ps}
  \caption{(a) The C~{\sc iv} absorption EW in troughs broader than 
1000~km~s$^{-1}$ (AI) plotted against \daox\ 
for high-ionization BAL and mini-BAL quasars. Mini-BAL quasars are plotted 
with red squares for \hbox{X-ray} detections and red stars for upper limits, while BAL 
quasars are plotted with black squares and arrows. Objects with large negative
(positive) \daox\ are \hbox{X-ray} weak (strong). From Gibson et~al. (2009b). 
(b) C~{\sc iv} maximum outflow velocity vs. 2500~\AA\ luminosity for 
high-ionization BAL quasars. The black squares (red squares) represent 
quasars with \daox\ above (below) the median \daox\ value. The solid
curve represents the ``velocity envelope'' from Figure~7 of
Ganguly et~al. (2007). Note that \hbox{X-ray} weaker sources (red squares) 
generally have higher outflow velocities than \hbox{X-ray} stronger 
sources (black squares). From Gibson et~al. (2009a).}
\end{figure}


\noindent
As mentioned above, \hbox{X-ray} absorption processes are likely critical to 
allowing wind driving and BAL formation, and thus one might expect to observe 
relations between the \hbox{X-ray} and UV absorption properties of BAL and
mini-BAL quasars. Evidence for such relations has been put forward
in the past (e.g., Laor \& Brandt 2002; Gallagher et~al. 2006). In
order to extend these studies, we have recently performed statistical 
analyses of BAL and mini-BAL quasars from the Sloan Digital Sky Survey (SDSS)
that also have (mostly serendipitous) sensitive \hbox{X-ray} coverage from 
\chandra\ and \xmm\ (Gibson et~al. 2009ab); this work has been enabled 
by the large numbers of SDSS BAL and mini-BAL quasars that have now
been cataloged (e.g., Gibson et~al. 2009a). Most of our analyses have utilized 
42 high-ionization BAL quasars, 
8 low-ionization BAL quasars, and  
48 high-ionization mini-BAL quasars 
(all are radio-quiet objects). 
These have been selected to have high-quality SDSS spectra and redshifts
of \hbox{$z=1.68$--2.28}, providing coverage of both 
C~{\sc iv} and Mg~{\sc ii} 
(the most common high-ionization and low-ionization BALs, respectively). 

Utilizing measurements of \aox\ and \daox\footnote{\aox\ is defined to be the 
slope of a power law connecting the rest-frame 2500~\AA\ and 2~keV monochromatic 
luminosities; i.e., $\alpha_{\rm ox}=0.3838 \log(L_{\rm 2~keV}/L_{2500~\mathring{\rm{A}}})$.
This quantity is well known to be correlated with $L_{2500~\mathring{\rm{A}}}$. We also define
$\Delta\alpha_{\rm ox}=\alpha_{\rm ox}({\rm Observed})-\alpha_{\rm ox}(L_{2500~\mathring{\rm{A}}})$, 
which quantifies the observed \hbox{X-ray} luminosity relative to that expected from the 
\aox-$L_{2500~\mathring{\rm{A}}}$ relation; \daox\ is useful for assessing the relative
level of \hbox{X-ray} weakness.}, we confirm past findings that
low-ionization BAL quasars are even \hbox{X-ray} weaker than high-ionization
BAL quasars, supporting the idea that remarkably heavy (and sometimes Compton-thick
with $N_{\rm H}\simgt 10^{24}$~cm$^{-2}$)
\hbox{X-ray} absorption is often present in quasars with low-ionization BALs. 
This remains the strongest connection found between \hbox{X-ray} and UV absorption 
properties for BAL quasars. We also find that the level of \hbox{X-ray} weakness for
mini-BAL quasars is intermediate between that of BAL and non-BAL quasars, 
supporting the idea that BAL and mini-BAL absorption are physically related
phenomena. 

We have found significant correlations, though with considerable 
scatter, between the level of \hbox{X-ray} weakness (\daox) and
(1) the C~{\sc iv} absorption EW in broad troughs (see Fig.~1a), 
(2) the maximum observed C~{\sc iv} outflow velocity ($v_{\rm max}$), and 
(3) the velocity width of C~{\sc iv} absorption. 
These correlations indicate that the strength of \hbox{X-ray} absorption is
indeed affecting the formation and acceleration of the UV outflow, with 
stronger and faster UV outflows generally requiring more X-ray
absorption. This is broadly consistent with the expectations of
models for BAL quasars that include a line-driven disk wind plus 
shielding gas.  
The above being said, however, we have identified a subset of mini-BAL quasars 
that have large $v_{\rm max}$ values \hbox{(15,000--30,000~km~s$^{-1}$)} yet show
no evidence for heavy \hbox{X-ray} absorption. Thus, while \hbox{X-ray} 
absorption generally promotes a strong and high-velocity UV outflow, it 
is perhaps not always required. Some other launching mechanism, perhaps 
magneto-rotational, could be responsible for accelerating the outflows of this 
subset of mini-BAL quasars. 

Our data also extend some of the findings of Laor \& Brandt (2002) and 
Ganguly et~al. (2007) on the ``envelope'' of $v_{\rm max}$ as a function of 
optical/UV luminosity, $L_{2500~\mathring{\rm{A}}}$ (see Fig.~1b). Specifically, 
BAL quasars with higher values of $v_{\rm max}$ (i.e., lying closer to the 
envelope) generally show stronger \hbox{X-ray} absorption. This extends the 
basic findings from Laor \& Brandt (2002), which were based upon Palomar-Green 
quasars, upward in luminosity by about an order of magnitude. Improved source 
statistics are now needed to explore the 
$v_{\rm max}$-$L_{2500~\mathring{\rm{A}}}$-\daox\ parameter space thoroughly, 
and these should be coming soon (see below).

\subsection{An X-ray Survey of Bright, Representative Radio-Loud BAL Quasars}

\begin{figure}
  \includegraphics[height=.35\textheight,angle=0]{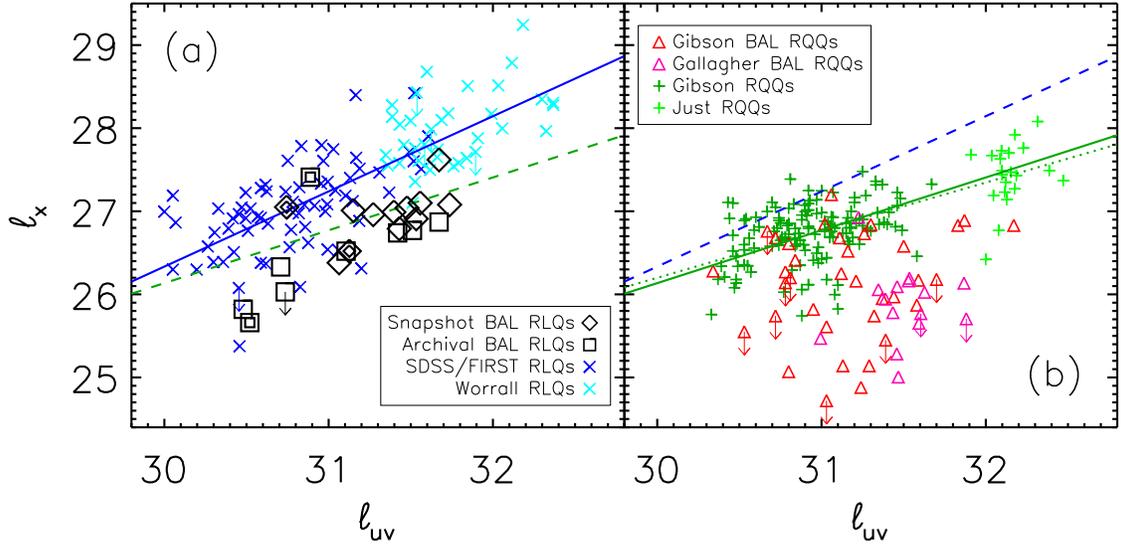}
  \caption{X-ray (2~keV) vs. optical/UV (2500~\AA) monochromatic 
luminosity (logarithmic) for BAL quasars as compared to non-BAL quasars. 
Panel (a) shows radio-loud BAL and non-BAL quasars (see symbol key). 
The solid and dotted lines in this panel show best-fit relations for 
non-BAL radio-loud and radio-quiet quasars, respectively. 
Panel (b) shows radio-quiet BAL and non-BAL quasars (see symbol key). 
The solid and dotted lines in this panel show best-fit relations for 
non-BAL radio-quiet quasars, while the dashed line shows the 
best-fit relation for non-BAL radio-loud quasars. 
Note, from comparing panels (a) and (b), that radio-loud BAL 
quasars are \hbox{X-ray} depressed relative to radio-loud non-BAL 
quasars, but not to the same degree as radio-quiet BAL quasars 
relative to radio-quiet non-BAL quasars. 
From Miller et~al. (2009); see this paper for further explanation.}
\end{figure}


\noindent
We have recently completed a systematic \chandra\ study of the \hbox{X-ray} properties
of 21 radio-loud BAL quasars (12 new and 9 archival observations;  
Miller et~al. 2009). Our new \chandra\ snapshot observations targeted the optically 
brightest radio-loud BAL quasars found in SDSS Data Release 3. A key 
goal of this work was to assess if radio-loud BAL quasars, which have both highly 
collimated relativistic jets as well as less-collimated winds, show similar
\hbox{X-ray} absorption properties to those of radio-quiet BAL quasars. 
Our sample includes objects with a variety of C~{\sc iv} absorption 
properties and is several times larger than those of past \hbox{X-ray} 
studies while also spanning a broader range of radio loudness 
(cf. Brotherton et al. 2005).

We find that the apparent \hbox{X-ray} luminosities of radio-loud BAL quasars are
depressed by factors of \hbox{$\approx 4$--9} compared to those of radio-loud 
non-BAL quasars that are matched in optical/UV luminosity (see Fig.~2a);
similar results are found when controlling for radio luminosity or 
both optical/UV and radio luminosity. This result is consistent with the 
presence of significant \hbox{X-ray} absorption in radio-loud BAL quasars. However, 
the degree of depression is typically less than that found when radio-quiet BAL 
quasars are compared to radio-quiet non-BAL quasars (see Fig.~2b). 
We also find a broad range of \hbox{X-ray} hardness ratios\footnote{The hardness
ratio is defined as $(H-S)/(H+S)$, where $H$ and $S$ are the hard-band \hbox{(2--8~keV)} 
and soft-band \hbox{(0.5--2~keV)} counts, respectively.} for radio-loud BAL
quasars, and there is no clear correlation between hardness ratio and 
the degree of \hbox{X-ray} weakness. Such a correlation is found for radio-quiet
BAL and mini-BAL quasars (e.g., Gallagher et~al. 2006; Gibson et~al. 2009b), as 
expected when absorption causes both spectral hardening and decreased apparent 
luminosity. 

We have used our constraints upon \hbox{X-ray} luminosity and spectral shape 
to examine quantitatively physical models for the \hbox{X-ray} emission of 
radio-loud BAL quasars. The most natural and straightforward interpretation of our
results is that radio-loud BAL quasars have both \hbox{X-ray} absorbers linked
with winds (as for radio-quiet BAL quasars) as well as sub-pc-scale \hbox{X-ray} 
emitting jets. Much, but likely not all, of the \hbox{X-ray} emission from 
the jet occurs on physical scales larger than that of the \hbox{X-ray} 
absorber. This unabsorbed jet-linked \hbox{X-ray} emission somewhat dilutes
the \hbox{X-ray} absorption signal relative to what is seen for radio-quiet
BAL quasars. Our basic model indicates that the \hbox{X-ray} absorber 
and the sub-pc-scale \hbox{X-ray} emitting jet must have roughly comparable 
sizes, and it is also broadly supported by our high-quality \hbox{X-ray} data 
on the bright radio-loud BAL quasar PG~1004+130 (Miller et~al. 2006).

\subsection{X-ray Iron~K BALs from Relativistic Outflows?}

\begin{figure}
  \includegraphics[height=.35\textheight,angle=0]{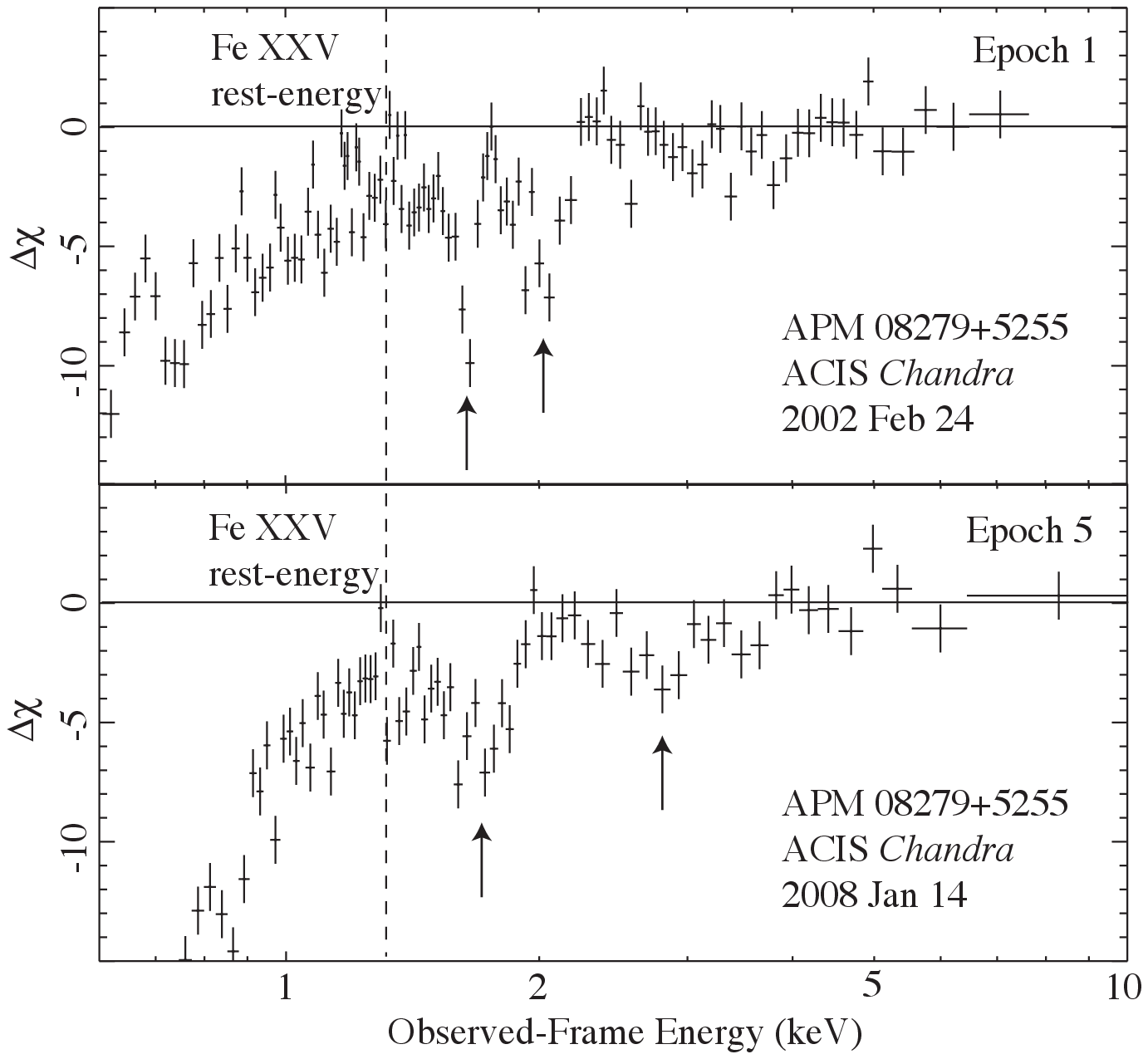}
  \includegraphics[height=.35\textheight,angle=0]{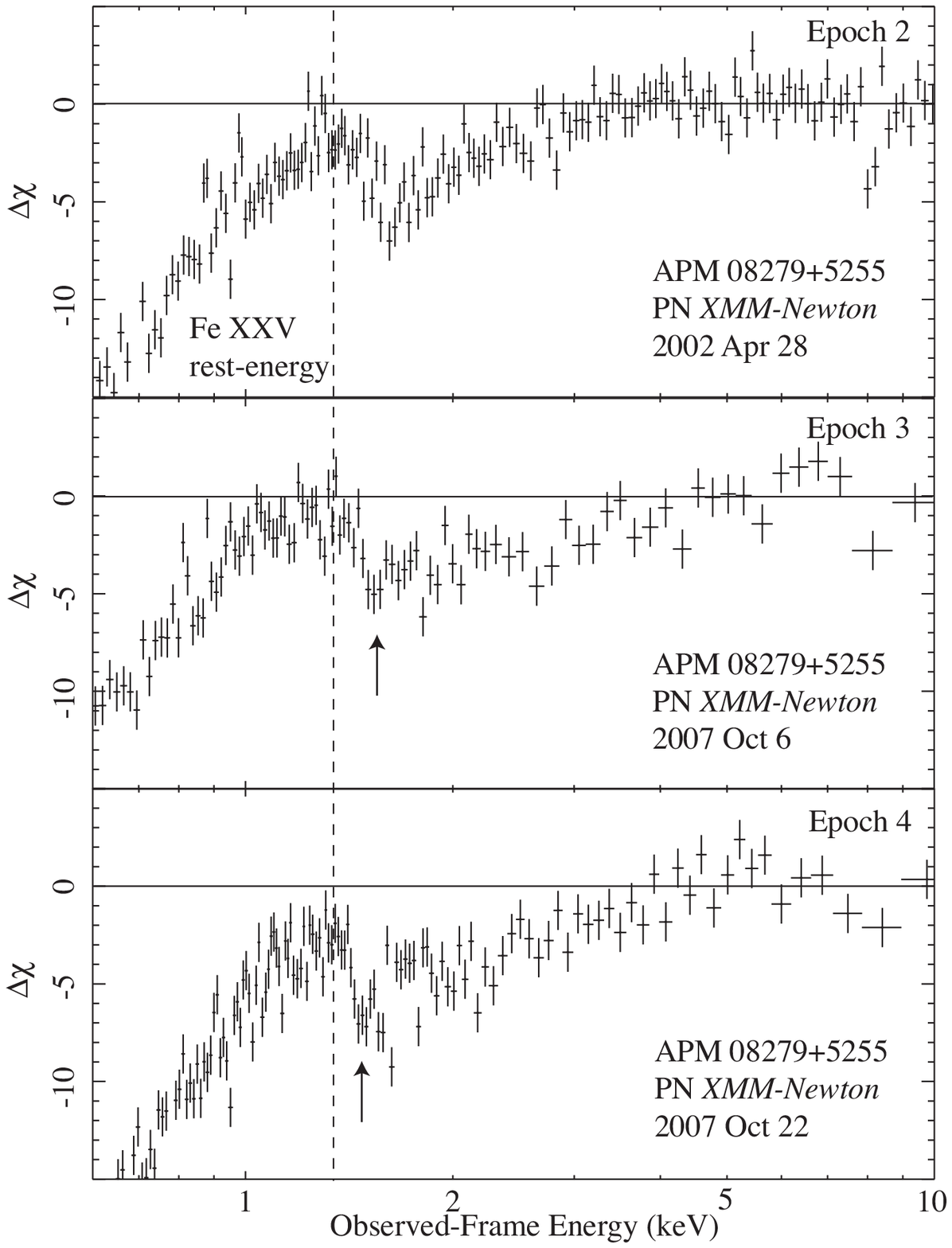}
  \caption{Residuals from fitting (a) \chandra\ and (b) \xmm\ spectra 
of the BAL quasar APM~08279+5255 with a power-law plus Galactic 
absorption model (data from \hbox{4.5--10~keV} are used in the fitting). 
The residuals are in units of sigma with error bars of size unity. 
Note that highly significant residuals from \hbox{1.5--3.5~keV}, 
corresponding to \hbox{7.5--17~keV} in the rest frame, are detected in all
spectra. From Chartas et~al. (2009); see this paper for further explanation.}
\end{figure}

\noindent
The kinetic luminosity of an AGN outflow, $L_{\rm kin}$, is proportional
to $f_{\rm c}\ r\ ({r\over \Delta r})\ N_{\rm H}\ v^3$, where 
$f_{\rm c}$ is its global covering factor, 
$r$ is its radius, 
$\Delta r$ is its thickness,
$N_{\rm H}$ is its column density, and 
$v$ is its velocity. 
Since the column density of the \hbox{X-ray} absorber in BAL quasars is 
generally large (\hbox{$N_{\rm H}\approx 10^{22}$--$10^{24}$~cm$^{-2}$}), 
this absorber could dominate the kinetic luminosity and 
mass-outflow rate {\it if\/} it is outflowing at about the velocity of 
the UV absorber (as is the case for local Seyfert galaxies) or higher. 
Unfortunately, however, the high-quality \hbox{X-ray} 
spectroscopic measurements of luminous quasars required for
assessment of absorption kinematics are challenging with current observatories. 
Therefore, the kinematic state of the \hbox{X-ray} absorber, and its 
relevance for AGN feedback into galaxies, is only now starting to 
become known for luminous quasars. 

In 2002, we proposed that a relativistic outflow was responsible for
creating remarkable \hbox{8--12~keV} (rest-frame) absorption features seen
in a \chandra\ spectrum of the luminous and gravitationally lensed 
BAL quasar APM~08279+5255 ($z=3.91$; Chartas et~al. 2002). These features were
best fit by absorption lines, so we proposed that they were 
\hbox{X-ray} BALs from iron~K transitions. The implied (projected)
velocity, even for highly ionized iron (the most conservative assumption), 
was remarkably large with \hbox{$v\approx 0.2$--$0.4c$}; this is much higher 
than the velocity of the UV BALs in this quasar. The implied
mass-outflow rate and kinetic luminosity are 
\hbox{$\sim 10$--30~$M_\odot$~yr$^{-1}$} and 
$L_{\rm kin}\sim 10^{47}$~erg~s$^{-1}$
for reasonable estimates of 
$f_{\rm c}$, $r$, $\Delta r$, and $N_{\rm H}$. 
While the theoretical mechanisms for launching such a powerful outflow
have not been worked out in detail, magneto-rotational or 
radiative processes could plausibly provide the requisite driving. 
The observed absorption features were detectable from APM~08279+5255 only 
due to its large gravitational lensing boost and high redshift, and such 
features could be present but undetected in the limited signal-to-noise 
spectra of many other distant BAL quasars. 

Since remarkable claims require remarkable evidence, we have been 
working to acquire additional spectra for APM~08279+5255 with one
primary goal being to test the model outlined in the previous paragraph. 
We now have 2 \chandra, 3 \xmm, and 3 \suzaku\ spectra of APM~08279+5255, 
and the relevant absorption features are consistently detected in all 8 
observations (see Fig.~3; Chartas et~al. 2009; Saez, Chartas, \& Brandt 2009). The 
features easily satisfy reasonable thresholds for significant detections, 
apparently unlike the case for some claimed relativistic narrow absorption 
lines (cf. Vaughan \& Uttley 2008).
Our multi-epoch observations also detect variability of the absorption-line
energies and equivalent widths; the full rest-frame energy range over which 
absorption has now been seen is \hbox{7.5--17~keV}. Such variability is 
expected given the likely small launching radius of the outflow, and it 
also helps to explain the somewhat different initial interpretations of 
the absorption features put forward by Chartas et~al. (2002) vs. 
Hasinger, Schartel, \& Komossa (2002). 

After our first report of iron~K BALs from APM~08279+5255, several
additional examples of possible relativistic outflows from luminous
quasars were reported (e.g., Chartas, Brandt, \& Gallagher 2003; 
Pounds et~al. 2003; Gibson et~al. 2005; Pounds \& Reeves 2009; 
Reeves et~al. 2009), some more convincing than others. Hopefully the 
number of luminous quasars with evidence for relativistic \hbox{X-ray} 
absorbing outflows will continue to grow, so that astronomers can build 
up a clearer picture of their kinetic luminosities and their apparently 
significant role in feedback. 

\section{Some Future Prospects}

The statistical \hbox{X-ray} studies of luminous quasar winds described above 
should advance rapidly over the coming decade, due to the combination of new 
large optical spectroscopic surveys of quasars and the continually
growing \hbox{X-ray} archives. For example, the \hbox{SDSS-III} BOSS survey 
(e.g., Schlegel, White, \& Eisenstein 2009) will obtain spectra for about
160,000 quasars at \hbox{$z\approx 2$--4}. These will generally have excellent 
coverage of transitions such as C~{\sc iv}, Si~{\sc iv}, and Al~{\sc iii}, 
and more than 30,000 new BAL and mini-BAL quasars should be discovered. 
Hundreds of these will already have sensitive archival \hbox{X-ray} 
coverage from \chandra\ and \xmm\ that may be used to define in detail
correlations between \hbox{X-ray} absorption, UV absorption, luminosity, 
and other physical properties. 

The \hbox{X-ray}  Microcalorimeter Spectrometer (XMS) on the 
{\it International X-ray Observatory (IXO)\/}
should dramatically advance studies of \hbox{X-ray} BALs in luminous
quasars. The large (up to \hbox{2--3~m$^2$}) \ixo\ collecting area 
will provide sufficient photon statistics for reliable \hbox{X-ray}  BAL detections 
in \hbox{20--80~ks} for many quasars. Equally important, the XMS
spectral resolution (2.5~eV) will allow study of the detailed 
structure of \hbox{X-ray} BALs, which is likely complex given the
known complexity of UV BALs. From these results, we expect 
substantial new insights into the energetics of quasar winds 
and their role in galaxy feedback. 


\begin{theacknowledgments}
We thank all of our collaborators on this work, including 
G.P. Garmire, 
M. Giustini, 
C. Saez, 
D.P. Schneider, and 
O. Shemmer. 
We acknowledge support from 
NASA LTSA grant NAG5-13035 (WNB, RRG, BPM), 
NASA grant SAO SV4-74018 (WNB, RRG, BPM), 
NASA grant NNX07AQ57G (WNB, RRG), and
the National Science and Engineering Research Council of Canada (SCG). 
\end{theacknowledgments}



\hyphenation{Post-Script Sprin-ger}
\begin{thebibliography}{0}
\expandafter\ifx\csname natexlab\endcsname\relax\def\natexlab#1{#1}\fi
\providecommand{\enquote}[1]{``#1''}
\expandafter\ifx\csname url\endcsname\relax
  \def\url#1{\texttt{#1}}\fi
\expandafter\ifx\csname urlprefix\endcsname\relax\def\urlprefix{URL }\fi
\providecommand{\eprint}[2][]{\url{#2}}

\bibitem[Chartas et~al. (2002)]{chartas:2002}
G. Chartas, W.N. Brandt, S.C. Gallagher, G.P. Garmire, 
\emph{ApJ} \textbf{579}, 169--175 (2002).
%

\bibitem[Chartas et~al. (2009)]{chartas:2009}
G. Chartas, C. Saez, W.N. Brandt, M. Giustini, G.P. Garmire, 
\emph{ApJ} in press (2009).
%

\bibitem[Ganguly et~al. (2007)]{ganguly:2007}
R. Ganguly, M.S. Brotherton, S. Cales, B. Scoggins, Z. Shang, M. Vestergard, 
\emph{ApJ} \textbf{665}, 990--1003 (2009a).
%

\bibitem[Gibson et~al. (2009a)]{gibson:2009a}
R.R. Gibson, et~al., 
\emph{ApJ} \textbf{692}, 758--777 (2009a).
%

\bibitem[Gibson et~al. (2009b)]{gibson:2009b}
R.R. Gibson, W.N. Brandt, S.C. Gallagher, D.P. Schneider, 
\emph{ApJ} \textbf{696}, 924--940 (2009b).
%

\bibitem[Miller et~al. (2009)]{miller:2009}
B.P. Miller, W.N. Brandt, R.R. Gibson, G.P. Garmire, O. Shemmer, 
\emph{ApJ} \textbf{702}, 911--928 (2009).
%

\bibitem[Saez et~al. (2009)]{saez:2009}
C. Saez, G. Chartas, W.N. Brandt, 
\emph{ApJ} \textbf{679}, 194--206 (2009).
%

\end{thebibliography}


\begin{thebibliography}{9}

\bibitem{brotherton:2006}
Brotherton M.S., Laurent-Muehleisen S.A., Becker R.H., Gregg M.D., 
Telis G., White R.L., Shang Z., 
\emph{AJ} \textbf{130}, 2006--2011 (2006).

\bibitem{chartas:2002}
G. Chartas, W.N. Brandt, S.C. Gallagher, G.P. Garmire, 
\emph{ApJ} \textbf{579}, 169--175 (2002).
%

\bibitem{chartas:2003}
G. Chartas, W.N. Brandt, S.C. Gallagher, 
\emph{ApJ} \textbf{595}, 85--93 (2003).
%

\bibitem{chartas:2009}
G. Chartas, C. Saez, W.N. Brandt, M. Giustini, G.P. Garmire, 
\emph{ApJ} in press (2009).
%

\bibitem{gallagher:2006}
S.C. Gallagher, W.N. Brandt, G. Chartas, R. Priddey, G.P. Garmire, R.M. Sambruna, 
\emph{ApJ} \textbf{644}, 709--724 (2006).

\bibitem{ganguly:2007}
R. Ganguly, M.S. Brotherton, S. Cales, B. Scoggins, Z. Shang, M. Vestergaard, 
\emph{ApJ} \textbf{665}, 990--1003 (2009a).
%

\bibitem{gibson:2005}
R.R. Gibson, H.L. Marshall, C.R. Canizares, J.C. Lee,  
\emph{ApJ} \textbf{627}, 83--96 (2005).
%

\bibitem{gibson:2009a}
R.R. Gibson, et~al., 
\emph{ApJ} \textbf{692}, 758--777 (2009a).
%

\bibitem{gibson:2009b}
R.R. Gibson, W.N. Brandt, S.C. Gallagher, D.P. Schneider, 
\emph{ApJ} \textbf{696}, 924--940 (2009b).
%

\bibitem{hasinger:2002}
G. Hasinger, N. Schartel, S. Komossa, 
\emph{ApJ} \textbf{573}, L77--L80 (2002).

\bibitem{laor:2002}
A. Laor, W.N. Brandt, 
\emph{ApJ} \textbf{569}, 641--654 (2002).

\bibitem{miller:2006}
B.P. Miller, W.N. Brandt, S.C. Gallagher, A. Laor, B.J. Wills, 
G.P. Garmire, D.P. Schneider, 
\emph{ApJ} \textbf{652}, 163--176 (2006).
%

\bibitem{miller:2009}
B.P. Miller, W.N. Brandt, R.R. Gibson, G.P. Garmire, O. Shemmer, 
\emph{ApJ} \textbf{702}, 911--928 (2009).
%

\bibitem{murray:1995}
N. Murray, J. Chiang, S.A. Grossman, G.M. Voit, 
\emph{ApJ} \textbf{451}, 498--509 (1995).

\bibitem{pounds:2003}
K.A. Pounds, J.N. Reeves, A.R. King, K.L. Page, P.T. O'Brien, M.J.L. Turner, 
\emph{MNRAS} \textbf{345}, 705--713 (2003).

\bibitem{pounds:2009}
K.A. Pounds, J.N. Reeves, 
\emph{MNRAS} \textbf{397}, 249--257 (2009).

\bibitem{proga:2004}
D. Proga, T.R. Kallman, 
\emph{ApJ} \textbf{616}, 688--695 (2004).

\bibitem{reeves:2009}
J.N. Reeves, et~al., 
\emph{ApJ} \textbf{701}, 493--507 (2009).

\bibitem{saez:2009}
C. Saez, G. Chartas, W.N. Brandt, 
\emph{ApJ} \textbf{679}, 194--206 (2009).
%

\bibitem{schlegel:2009}
D. Schlegel, M. White, D. Eisenstein, 
\emph{arXiv} 0902.4680 (2009).

\bibitem{vaughan:2008}
S. Vaughan, P. Uttley, 
\emph{MNRAS} \textbf{390}, 421--428 (2008).

\end{thebibliography}
\end{document}